\documentclass[a4paper,twoside]{article}

\usepackage{epsfig}
\usepackage{subcaption}
\usepackage{calc}
\usepackage{amssymb}
\usepackage{amstext}
\usepackage{amsmath}
\usepackage{amsthm}
\usepackage{multicol}
\usepackage{pslatex}
\usepackage{apalike}
\usepackage{algorithm2e}
\usepackage{url}
\usepackage[bottom]{footmisc}
\usepackage[table]{xcolor} 
\usepackage{SCITEPRESS}   

\usepackage{hyperref}
\hypersetup{hypertex=true,
            colorlinks=true,
            linkcolor=black,
            anchorcolor=black,
            citecolor=black,
            urlcolor=black}

\usepackage[shortcuts,acronym]{glossaries}
\newacronym{ukf}{UKF}{Unscented Kalman Filter}
\newacronym{ut}{UT}{Unscented Transformation}
\newacronym{dos}{DoS}{{Denial of Service}}
\newacronym{ml}{ML}{{Machine Learning}}
\newacronym{ad}{AD}{{Attack Detection}}
\newacronym{mlp}{MLP}{{multi-layer perceptron}}
\newacronym{lstm}{LSTM}{{Long Short-Term Memory}}
\newacronym{cps}{CPS}{{Cyber–Physical System}}
\newacronym{adas}{ADAS}{{Advanced Driver Assistance Systems}}
\newacronym{cusum}{CUSUM}{{Cumulative Sum}}
\newacronym{add}{AD}{{Attack Detection}}
\newacronym{relu}{ReLU}{{Rectified Linear Unit}}
\newacronym{avs}{AVs}{Autonomous Vehicles}
\newacronym{cavs}{CAVs}{Connected Autonomous Vehicles}
\newacronym{pwm}{PWM}{Pulse Width Modulation}
\newacronym{can}{CAN}{Controller Area Network}
\newacronym{v2v}{V2V}{Vehicle-to-Vehicle}
\newacronym{cnn}{CNN}{Convolutional Neural Network}
\newacronym{fdi}{FDI}{False Data Injection}

\begin{document}

\title{Data-Driven Intrusion Detection in Vehicles: Integrating \ac{ukf} with Machine Learning \thanks{Accepted in Proceedings of the 21st International Conference on Informatics in Control, Automation and Robotics (ICINCO 2024).}}

\author{\authorname{Shuhao Bian\sup{1}, Milad Farsi\sup{1}, Nasser L. Azad\sup{1}, Chris Hobbs\sup{1} 
}
\affiliation{\sup{1}Systems Design Engineering Dep. , University of Waterloo, 200 University Ave W, Waterloo, Canada}
\email{\{s6bian, mfarsi, nlashgarianazad\}@uwaterloo.ca, cwlh@farmhall.ca}
}


\abstract{\glsresetall In the realm of \ac{cps}, accurately identifying attacks without detailed knowledge of the system's parameters remains a major challenge. When it comes to \ac{adas}, identifying the parameters of vehicle dynamics could be impractical or prohibitively costly.
To tackle this challenge, we propose a novel framework for attack detection in vehicles that effectively addresses the uncertainty in their dynamics. Our method integrates the widely used \ac{ukf}, a well-known technique for nonlinear state estimation in dynamic systems, with machine learning algorithms. This combination eliminates the requirement for precise vehicle modelling in the detection process, enhancing the system's adaptability and accuracy.
To validate the efficacy and practicality of our proposed framework, we conducted extensive comparative simulations by introducing \ac{dos} attacks on the vehicle systems' sensors and actuators.}

\onecolumn \maketitle \normalsize \setcounter{footnote}{0} \vfill

\section{{INTRODUCTION}}
\label{sec:introduction}
\glsresetall
\ac{cps} refers to the integration of computation with physical processes. Therefore, cyber attacks on these systems can cause severe consequences. The reliable operation of \ac{adas} depends on the accurate functioning of various sensors and power management systems. If these elements are targeted by malicious attackers, passengers, pedestrians and drivers could be exposed to significant safety risks, potentially endangering their lives. This situation underscores the critical necessity for attack detection methods within autonomous driving systems. By identifying potential threats as they occur, the system can initiate appropriate protective actions to safeguard passengers, pedestrians and drivers.

\ac{dos} attack is one of the most well-known cyber attacks, and it has become more prevalent since 2004 \cite{mirkovic2004taxonomy}. These attacks purposefully flood networks with too much traffic, overwhelming systems and compromising service availability. \ac{dos} attacks can cause significant operational disruptions in \ac{avs} and \ac{cavs}, which rely heavily on continuous and secure communication channels for services such as navigation, real-time traffic updates, and vehicle-to-vehicle (V2V) communication. The absence of connectivity not only affects the safety elements essential to AV operations but also degrades the system's ability to make intelligent decisions on the road. Multiple recent studies can be found in the literature, providing insight into the different classes of attacks and defense mechanisms developed, such as \cite{naqvi2022systematic,marcillo2022security,al2019intrusion,banafshehvaragh2023intrusion,icinco23}.

Concerning vehicle security, there exist various applications of machine learning techniques in the literature for detecting anomalies in different networks. One group of approaches considers the possibility of attacks on communications between vehicles themselves and between vehicles and roadside infrastructure. \cite{canh2023machine} seeks to build and evaluate a particular attack detection system that employs four specific discriminating features. A collected dataset is then utilized to train and evaluate several machine learning and statistical models, allowing for a comparative examination of their efficacy. The suggested strategy focuses on early detection, allowing for timely and effective countermeasures.

Another class of intrusion detection has focused on in-vehicle communications used for exchanging data between different control units of vehicles. \cite{berger2018comparative} evaluates various machine learning methods, including deep learning, for in-vehicle intrusion detection systems. In a more recent technique, \cite{aldhyani2022attacks} implemented deep learning approaches like \ac{cnn}s and \ac{cnn}-\ac{lstm} hybrid models to detect attacks such as spoofing, flooding, and replay attacks on the \ac{can} bus. Other similar techniques can be found in \cite{pawar2022cyber}. A comprehensive survey of the techniques presented, in the literature, can be found in \cite{rajapaksha2023ai}.

The majority of these approaches focus on network traffic and data package analysis to discover recurring patterns in normal operation during the training stage, and then use them to detect anomalies during the exploitation stage. Therefore, the dynamic aspects of different subsystems in the vehicles receive less attention. On the other hand, dynamic models are the foundation of vehicle design, allowing engineers to predict and optimize the performance of numerous vehicle systems under a variety of operating situations. Exploiting such models for intrusion detection allows us to identify anomalies by continuously comparing real-time data against the predicted normal behavior. Through this proactive strategy, intrusions can be detected early and promptly addressed to reduce potential damage. \cite{ju2022survey} provides a review of such techniques from a control perspective. In line with these approaches, in this study, we employ a rather system dynamics model-based technique as a defense mechanism. Moreover, since the parameters of the vehicle, such as weight, tire conditions, and engine characteristics, can change over time or under different conditions, we have found machine learning techniques particularly advantageous as they offer high adaptability. 


The \ac{ukf} has been employed successfully in various fields of applications, such as power and automotive systems. In power systems, the \ac{ukf} contributes to stability and operational integrity by accurately evaluating the state of electrical grids and identifying potential interruptions \cite{du2022review,rashed2022false}. The \ac{ukf} has played an important role in improving cyber threat detection in-vehicle systems \cite{zhang2021cyber}, particularly \ac{avs} and \ac{cavs}. The \ac{ukf} improves the system's robustness against \ac{dos} attacks by enabling real-time, precise anomaly detection, preventing possible threats from causing harm. In \cite{vzivkovic2018detection}, the authors employed the \ac{ukf} to predict and update state variables from a previously known state to detect \ac{fdi} attacks. They compared the results to those obtained using a common weighted least squares-based state estimation technique. They observed that the state variables under attack significantly deviated between them, which can be used to detect the attack. 

 In this paper, we describe a novel approach for developing an attack detection system tailored to vehicles. Our proposed approach integrates the \ac{ukf} with a learning-based module to obtain a resilient adaptive framework. This feature eliminates the requirement for detailed vehicle modeling in the attack detection process, simplifying implementation while retaining accuracy. The framework's effectiveness is strengthened by the use of a \gls{cusum} algorithm with a sliding window for responsive anomaly detection, and by incorporating the learned dynamics to predict and compare real-time data against expected behavior. By leveraging the \ac{ukf}'s capabilities in handling non-linear dynamics, our proposed algorithm significantly improves the robustness and accuracy of intrusion detection in cyber-physical systems, particularly in \ac{adas}. We conducted extensive simulations using CARLA, a simulation platform commonly used in autonomous driving research, to evaluate the efficacy and feasibility of our system. This made it possible for us thoroughly to test our structure in practical settings and make sure it can operate dependably in actual situations.

 The rest of the paper is organized in the following order: In Section~\ref{problem_formulation}, we formulate the problem by assuming a \ac{dos} attack on the actuator. Section~\ref{main_results} presents the attack detection framework, introducing the main algorithm. In the following Section~\ref{sim_results}, we discuss the detailed simulation results. 
\begin{figure*}[h]
  \centering
   {\epsfig{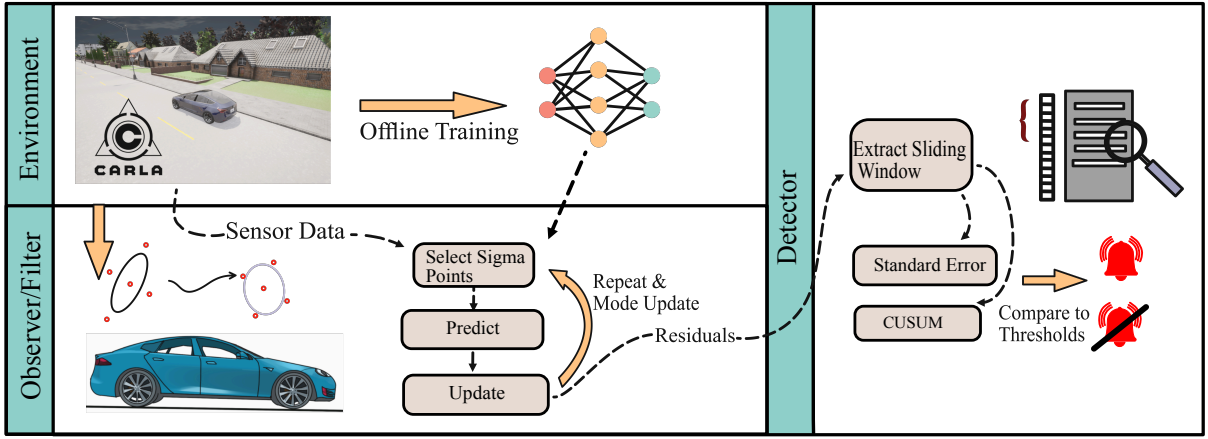}}
  \caption{The flow chart of the algorithm.}
  \label{fig:flow}
 \end{figure*}
\section{\uppercase{Problem Formulation}}

\label{problem_formulation}
The system model is described by the following equations
\begin{equation}
\begin{aligned} \label{objective}
    x^{k+1}&=f(x^{k},u^{k}),\\
    y^k&=h(x^{k}),
\end{aligned}   
\end{equation}
where $x^k\in \mathbf{R}^{n}$, represents the state space of the vehicle, and $u^k\in \mathbf{R}^{p}$, denotes the control input. $y^k\in \mathbf{R}^{m}$ is the measurement vector. Moreover, \( f: \mathbf{R}^n \times \mathbf{R}^p \rightarrow \mathbf{R}^n \) represents the state transition function that defines the next state \( x^{k+1} \) based on the current state \( x^k \) and the control input \( u^k \). The function \( h: \mathbf{R}^n \rightarrow \mathbf{R}^m \) denotes the measurement function that maps the current state \( x^k \) to the measurement vector \( y^k \). $k$ gives the time step. Next, we will model the \ac{dos} attack on various elements of the control loop.

\subsection{Attack Model}
Considering that the attacks are implemented in the cyber layer, we construct the cyber attack in a discrete space, as shown below.

\subsubsection{Actuator Attack}
To model the effect of the attack, the dynamics in (\ref{objective}) are modified as 
\begin{align} \label{actuator attack formula}
    x^{k+1}=\begin{cases}
     f(x^{k},a^{k}), &k\in \alpha,\\
    f(x^{k},u^{k}), &k \notin \alpha,\\
    \end{cases}
\end{align}
where we denote the set of time steps during which the attack is active using $\alpha$. Moreover, we assume that \ac{dos} can arbitrarily affect a subset of components of $u^k$. This can be further described by
\begin{equation}
a^{k}[i] = 
\begin{cases}
0, & i \in \Gamma_a, \\
u^{k}[i], & \text{otherwise},
\end{cases}
\end{equation}
where $\Gamma_a$ denotes the set of indices corresponding to the actuators that are affected by the attack, and $i$ ranges over the set $\Gamma_a \subseteq\{1, 2, \ldots, p\}$. Accordingly, a particular attack strategy can be represented by choosing nonempty sets $\alpha$ and $\Gamma_a$, which specify the time of the attack and the indices of targeted actuators, respectively. 

\subsubsection{Sensor Attack}
In this section, we define the sensor attack in a similar fashion. We modify the measurements relation given in equation (\ref{objective}) to reflect the \ac{dos} attack on some specific set of sensors. The equation
\begin{equation}
y^{k} =
\begin{cases}
y_a^{k}, & k\in \beta,\\
h(x^k), & k \notin \beta, \\
\end{cases}
\end{equation}
switches the measurements to $y_a^{k}$ that may be manipulated by the anomaly, at different time steps specified by set $\beta$. Moreover, based on a particular attack pattern, a list of targeted sensors is given by $\Gamma_s$, with the attacked  components being set to zero. This is shown below
\begin{equation}\label{eq: longitudinal dynamics}
y_a^{k}[i] = 
\begin{cases}
0, &  i \in \Gamma_s, \\
y^{k}[i], & \text{otherwise},
\end{cases}
\end{equation}
where $\Gamma_s \subseteq\{1, 2, \ldots, m\}$ denotes the set of indices corresponding to the components of $y^{k}$ that are blocked by the attacker. Here, $m$ refers to the dimension of the measurement vector.

\subsection{Vehicle Model}
In order to demonstrate the performance of the developed approach, we use the Kalman filter as a baseline approach for comparison in Section \ref{sim_results}. It should be noted that we aim to obtain a sample-based technique, meaning that we treat the dynamics as a black box without using the analytical model. However, since the dynamic model are essential for running the Kalman Filter, we concisely present the equations of the vehicle in what follows.

According to \cite{Takahama201820184095}, a vehicle dynamics model can be obtained as below. The longitudinal dynamics of the vehicle is given by
\begin{equation}\label{eq: rtravel}
M_{vehicle}\dot{v}_h=M_{vehicle}a_f-r_{travel},
\end{equation}
where $M_{vehicle}$ is the mass of the vehicle, $a_f$ is the traction force converted to acceleration and $r_{travel}$ is the travel resistance. The model of the $r_{travel}$ can be described as
\begin{equation}\label{vehicle force}
r_{travel}=r_{air}(v_h^2)+r_{roll}(v_h)+r_{accel}(\dot{v}_h)+r_{grad}(\theta),
\end{equation} 
where $ r_{air} = \frac{1}{2} \rho C_d A v_h^2 $ is the air drag, $\rho$ is the air density, $C_d$ is the drag coefficient, $A$ is the frontal area of the vehicle, and $v_h$ is the vehicle speed. The rolling resistance is given by $ r_{roll} = C_r M_{vehicle} g v_h $, where $C_r$ and $g$ denote the rolling resistance coefficient and the acceleration due to gravity, respectively.

Then, the acceleration resistance is $ r_{accel} = M_{vehicle} \dot{v}_h $, where $\dot{v}_h$ is the acceleration of the vehicle. Finally, $ r_{grad} = M_{vehicle} g \sin(\theta) $, where $\theta$ is the slope angle of the road.

\begin{algorithm}[h]
\caption{\ac{cusum} Algorithm with Sliding Window for Attack Detection}\label{alg:cusum_sliding_window}
\KwData{Sensor signal, Input signal}
\KwResult{Attack State: $f_{ad}$ (0 for no attack, 1 for attack detected)}
\textit{Initialisation:} \; 
Initialize the cumulative sum, $s_1,s_2 \gets 0$ \;
Initialize the attack detection flag, $f_{ad}\gets 0$\;
Initialize the threshold value, $\textit{threshold}$ \;
Define the window size $N$ \;
Initialize an empty list $Q$, to store the last $N$ residuals \;
\textit{Detection Loop:} \;
\While{each new sensor signal input $y$}{
    Perform the selecting sigma points step\;
    Perform the prediction step of UKF to estimate the next state \;
    Perform the update step of UKF with signal $y$ to obtain residual $r$ \;
    Append $r$ to $Q$ \;
    Update the UKF model \tcp*[f]{using equation (\ref{alg:learning rate})} \;
    \If{Size of $Q > N$}{
        Remove the oldest residual from $Q$ \;
    }
    Update cumulative sum using residuals within $Q$: $s_1 \gets \sum Q$ \tcp*[f]{using equation (\ref{alg:residual_test1})} \;
    Update standard error using residuals within $Q$: $s_2 \gets std(Q)$ \tcp*[f]{using equation (\ref{alg:residual_test2})} \;
    \eIf{$abs(s_1) > \textit{threshold}_1$ \textbf{and} $s_2 > \textit{threshold}_2$}{
        $f_{ad}\gets 1$ \tcp*[f]{attack detected immediately upon detection} \;
    }
    {
        $f_{ad}\gets 0$\;
        Continue monitoring \;
    }
}
\end{algorithm}

\section{Attack Detection Framework}
\label{main_results}
This section presents the required components to construct the attack detection framework.

\subsection{Unscented Kalman Filter}
In the \ac{ukf}, knowing the detailed model is unnecessary. The \ac{ukf} uses sigma points to sample the input and obtain the corresponding output. It is a method similar to the Monte Carlo approach but requires only a tiny number of sigma points. 
Next, we will briefly introduce the \ac{ukf}.

\subsubsection{\ac{ut}}

As described by \cite{wan2000unscented}, the \ac{ut} is a technique developed for the generation of sigma points that are capable of undergoing nonlinear transformations expressed as $f(x^k, u^k)$. This approach is particularly valuable when dealing with a multitude of random vectors, each residing in an $n$-dimensional space ($x^k\in \mathbf{R}^n$), characterized by a mean $\Bar{x}^k$ and a covariance matrix $P^k$.

\begin{figure}[h]
    \centering
    \includegraphics[clip, trim=0.0cm 0.0cm 0.0cm 0.0cm, width=1\linewidth]{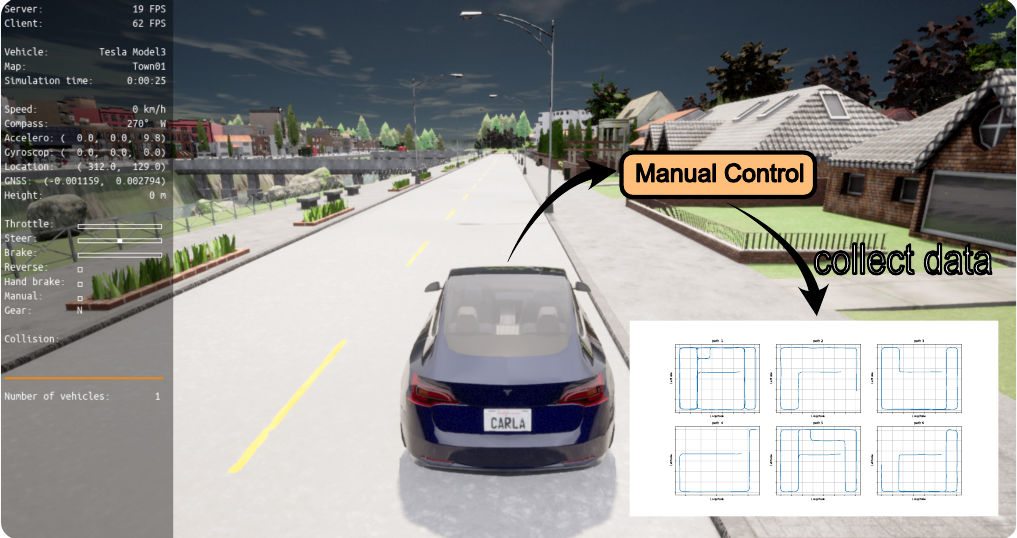}
    \caption{CARLA Environment}
    \label{fig:CARLA}
\end{figure}

\subsubsection{Select Sigma Points}

The sigma points $\chi_i^k$ are chosen to capture the true mean and covariance of the random variable $x^k$, enabling an accurate propagation through the nonlinear function $f(x^k, u^k)$. This selection ensures that the mean and covariance of the sigma points match $\Bar{x}^k$ and $P^k$, respectively, providing an effective mechanism for estimating the statistical properties of $x^{k+1}$. The sigma points are selected according to

\begin{equation}\label{unscented transform}
\left\{
\begin{array}{lll}
\chi_0^k&=\Bar{x}^k,&\\
\chi_i^k&=\Bar{x}^k+(\sqrt{(n+\lambda)P^k})_i&i=1,...,n,\\
\chi_i^k&=\Bar{x}^k-(\sqrt{(n+\lambda)P^k})_i&i=n+1,...,2n,\\
W_0^{(m)}&=\frac{\lambda}{n+\lambda},&\\
W_0^{(c)}&=\frac{\lambda}{n+\lambda}+(1-\phi^2+\beta),&\\
W_i^{(m)}&=W_i^{(c)}=\frac{1}{2(n+\lambda)}&i=1,...,2n.\\
\end{array}
\right.
\end{equation}

According to \cite{wang2023combining}, $\lambda$ is a key parameter calculated using $\lambda = \phi^2(n+\kappa)-n$, $\phi$ determines the dispersion degree of $\sigma$ points, and $\kappa$ is typically $3-n$. $\beta$ is for integrating prior knowledge about $x$'s distribution.

\subsubsection{Predict}

The predicted state and covariance are
\begin{equation}
\begin{aligned} \label{predicted state and covariance}
    \bar{x}^{k+1|k}&=\sum_{i=0}^{2n}W_i^{(m)}\chi_i^{k+1|k},\\
    P^{k+1|k}&=\sum_{i=0}^{2n}W_i^{(c)}\left(\chi_i^{k+1|k}-\bar{x}^{k+1|k}\right)\left(\chi_i^{k+1|k}-\bar{x}^{k+1|k}\right)^T,\\
    \bar{y}^{k+1|k}&=\sum_{i=0}^{2n}W_i^{(m)}\mathcal{Y}_i^{k+1|k},\\
    P_{yy}^{k+1}&=\sum^{2n}_{i=0}W_i^{(c)}\left(\mathcal{Y}_i^{k+1|k}-\bar{y}^{k+1|k}\right)\left(\mathcal{Y}_i^{k+1|k}-\bar{y}^{k+1|k}\right)^T,\\
    P_{xy}^{k+1}&=\sum^{2n}_{i=0}W_i^{(c)}\left(\chi_i^{k+1|k}-\bar{x}^{k+1|k}\right)\left(\mathcal{Y}_i^{k+1|k}-\bar{y}^{k+1|k}\right)^T,\\
\end{aligned}
\end{equation}
where $\chi^{k+1|k}_i=f(\chi^{k}_i,u^{k})$, $\mathcal{Y}_i^{k+1|k}=h(\chi^{k}_i,u^{k})$, for $i=0,...,2n$.

\subsubsection{Update}

The update step can be defined as
\begin{equation}\label{update part}
    \begin{aligned}
    K^{k+1}&=P_{xy}^{k+1}\left(P^{k+1}_{yy}\right)^{-1},\\
    r^{k+1}&=y^{k+1}-\bar{y}^{k+1|k},\\
    x^{k+1}&=\bar{x}^{k+1|k}+K^{k+1} r^{k+1},\\
    P^{k+1}&=P^{k+1|k}-K^{k+1}{P_{xy}^{k+1}}^T,
    \end{aligned}
\end{equation}
where $K$ is the Kalman gain, $r$ is the residual. These values are updated when sensors provide new measurements.

\subsection{Machine Learning}
Obtaining the parameters of a vehicle might often be infeasible or economically prohibitive. In the update phase of the \ac{ukf}, we employ \ac{ml} techniques to predict the vehicle's subsequent state.

For this purpose, we selected the \ac{mlp} network. This neural network receives control commands ($u$) --- encompassing the throttle-brake and steering angle --- as well as the vehicle's current state, which includes velocity, angular velocity, and acceleration, as its input. The network is designed to output the vehicle's acceleration for the next time step. Designed to predict the vehicle's acceleration at the next time step, it was necessary to account for the delay between the input commands and the resultant state changes. To address this, according to \cite{xu2019automated}, we selected a model output that reflects the vehicle's acceleration 50 milliseconds after the input, ensuring the training data adequately captures the dynamics of the system.

The \ac{mlp} is structured with three distinct layers: an input layer comprising 5 units, a hidden layer containing 20 units, and a single-unit output layer. The activation function utilized within the hidden layer is the \ac{relu}.

With this predicted acceleration, we can determine the vehicle's upcoming velocity using Newton's second law, thereby integrating \ac{ml} predictions seamlessly into the \ac{ukf}'s update mechanism for enhanced estimation accuracy.

\subsection{Model update methods}

In order to make the detector adaptive, we deployed an algorithm that can make the detector update automatically. In the proposed detector, we updated the model of the vehicle when it got new sensor signals. To make sure the model is not updated when the attacker implements the \ac{dos} attack on the car model and is updated when the car is running normally, we update the learning rate according to 
\begin{align} \label{alg:learning rate}
    l = 1-\frac{1}{1+e^{-S_{rate} \|r\|_2}}.
\end{align}
Here, $ l $ denotes the learning rate, $ r $ represents the residual of the car, $ S_{rate} $ is the scale of the residual, and $\|\cdot\|_2$ denotes the $L_2$ norm of a vector.

\subsection{Attack Detection Method}

As a widely used attack detector in many literature \cite{liu2019secure}, \ac{cusum} is selected as the detector. In \cite{van2012sensor}, they also calculated a moving window average for the residual. In practice, there is often a gap between theoretical models and engineering applications. Therefore, we introduce $W_{r1}$ and $W_{r2}$ to fine-tune the target parameters. Additionally, we used two tests to detect whether the target is under attack.

Test~1 is shown by
\begin{align} \label{alg:residual_test1}
    s_1^k = \sum ^{k}_{i=k-N+1}W_{r1} r^i,
\end{align}
where $r$ is the residual updated by the \ac{ukf}, $W_{r1}^T\in \mathbf{R}^{m}$, $N$ is the length of the sliding window.

Test~2 is described as
\begin{align} \label{alg:residual_test2}
    s_2^k = \sum ^{k}_{i=k-N+1} (r^i-\bar{r})W_{r2}(r^i-\bar{r})^T,
\end{align}
where $\bar{r}$ is the average of the sliding window for the residuals, $W_{r2}\in \mathbf{R}^{m\times m}$ is the weight of the residual.

Then $s_1$ and $s_2$ will be compared with two thresholds, $t_1$ and $t_2$, respectively. Then the alarm is triggered according to
\begin{equation} \label{trig_alarm}
    A = \begin{cases} 
    S_p, &  s_1 > t_1 \text{ and } s_2 > t_2, \\
    S_n, &  s_1 \leq t_1 \text{ or } s_2 \leq t_2.
    \end{cases}
\end{equation}

As illustrated in Figure \ref{fig:flow}, the trained model derived from historical data is subsequently integrated into the \ac{ukf}. The mechanism of the detector is detailed in Algorithm~ \ref{alg:cusum_sliding_window}.

\section{Simulation Results}
\label{sim_results}

To demonstrate the efficacy of the proposed approach in \ac{dos} attacks, we chose CARLA \cite{dosovitskiy2017carla} as our simulator, which can simulate real-world dynamics and generate sensor signals in real time. This allows for accurate and timely responses in the simulation environment. Moreover, to highlight the superiority of the proposed method, we compare it with the traditional Kalman filter.

All neural network training was conducted in Python on the Ubuntu operating system. The hardware configuration used included an AMD Ryzen 9 processor with 16 cores, clocked at 3.40 GHz, and 64GB of RAM. The Kalman filter and \ac{ukf} were implemented using the FilterPy library \cite{labbe2024filterpy}, while the neural network was developed with PyTorch \cite{paszke2019pytorch}.

In the following subsection, we detail the data preprocessing methods, the implementation of attack scenarios, the simulation parameters, and the resulting simulation outcomes.

\subsection{Data Preprocessing}
Both manual driving data and autonomous driving data generate command signals (throttle, brake, and steering angle), which can be used as training data. To obtain representative data, we used manual control to generate data in the CARLA environment, as shown in Figure~\ref{fig:CARLA}. 

To achieve better training results, we need to ensure that the data are valid by removing all outliers.

The brake and throttle inputs are combined and normalized into a unified control signal ranging from 0~to~1 according to 
\begin{equation}
\label{eq:brake throttle}
u = \frac{T - B +1}{2},
\end{equation}
where $u$ is the unified control signal, $T\in[0,1]$ is the throttle input, $B\in[0,1]$ is the brake input, $T_{max}$ is the maximum throttle value and $B_{min}$ is the minimum brake value. When $u = 0$, it represents full braking, and when $u = 1$, it represents full throttle.

\subsection{Attack Implementation}

We implemented the attack at 20 seconds in the simulation. To validate the algorithm, we chose a \ac{dos} attack signal in the form of a \ac{pwm} structure. When the signal is 1, it indicates that the monitor is blocked.

\subsection{Parameters}

To make a comparison, we chose a typical Kalman filter. The parameters of equation \eqref{vehicle force} are shown as table~\ref{tab:kf parameters}.

\begin{table}[h] 
    \centering
    \caption{Parameters of Kalman filter}
    \label{tab:kf parameters}
    \begin{tabular}{|c|c|c|} 
        \hline
        \textbf{Name} & \textbf{Value}\\ \hline
        $r_{air}$ & 68.9 N\\ \hline
        $r_{roll}$ & 271.6 N\\ \hline
        $r_{grad}$ & 0 N\\ \hline
    \end{tabular}
\end{table}

If there is no slope, $r_{grad}=0$. Therefore, the total travel resistance $r_{travel}$ is given by $r_{travel}=r_{air}+r_{roll}+r_{grad}=340.5N$. In order to make the detector more accurate, we have $W_{r1}=\begin{bmatrix}1 & 0.01 & 0\end{bmatrix}^T$, $W_{r2}=\begin{bmatrix}
1 & 0 & 0 \\
0 & 0.01 & 0 \\
0 & 0 & 0 \\
\end{bmatrix}
$ and we focus on the acceleration residual to detect the attack.

The learning parameters are detailed in Table~\ref{tab:learning parameters}. The optimizer used is Adam \cite{kingma2014adam}.

\begin{table}[h] 
\vspace{-0.2cm}
    \caption{Parameters of Machine Learning}  \centering
    \label{tab:learning parameters}
    \begin{tabular}{|c|c|c|} 
        \hline
        \textbf{Name} & \textbf{Value}\\ \hline
        Learning rate & 0.001\\ \hline
        Train size/Total & 0.8\\ \hline
        Epochs & 1000\\ \hline
        \ac{mlp} batch size & 64\\ \hline
        \ac{mlp} epochs & 1000 \\ \hline
    \end{tabular}
\end{table}

To improve the clarity of the results, we simplified the model; the value of $t_2$ is based on $t_1$, maintaining a fixed proportional relationship as shown by
\begin{equation} \label{t1_t2}
t_2 = \gamma t_1,
\end{equation}
where $\gamma$ is a tuning parameter. In this simulation, $\gamma$ is set to 0.04.

\subsection{Results}

In this study, the performance of the Kalman filter and \ac{ukf} for detecting attacks in autonomous vehicles was evaluated. The experiments were conducted under two scenarios: actuator attack and sensor attack.

\begin{figure*}
  \centering
  \includegraphics[width=\textwidth, height=18.5cm]{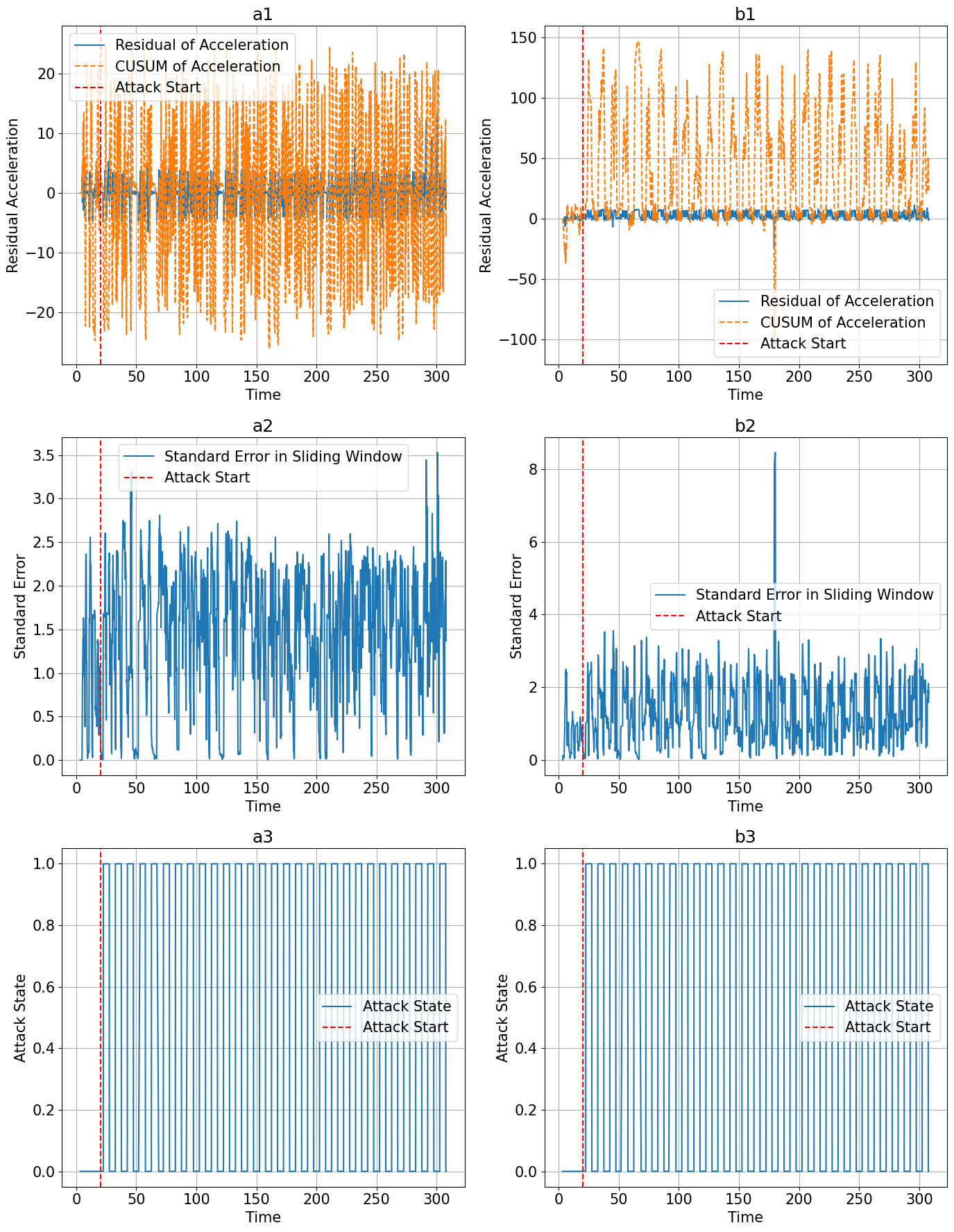}
  \caption{The results of the attack detector are as follows: for the Kalman filter, $a1$ represents the residual and \ac{cusum} of residuals in the sliding window, $a2$ denotes the standard error of residuals in the sliding window, and $a3$ indicates the attack state; for the proposed method, $b1$ signifies the residual and \ac{cusum} of residuals in the sliding window, $b2$ refers to the standard error of residuals in the sliding window, and $b3$ denotes the attack state of the \ac{ukf}.}
  \label{fig:kf and ukf}
\end{figure*}

\subsubsection{Actuator Attack}

To validate the detector, we measure its performance using four metrics: false positive alarm rate, true positive alarm rate, false negative alarm rate, and true negative alarm rate.

\begin{figure}[h]
    \centering
    \includegraphics[clip, trim=0.0cm 0.0cm 0.0cm 0.0cm, width=1\linewidth]{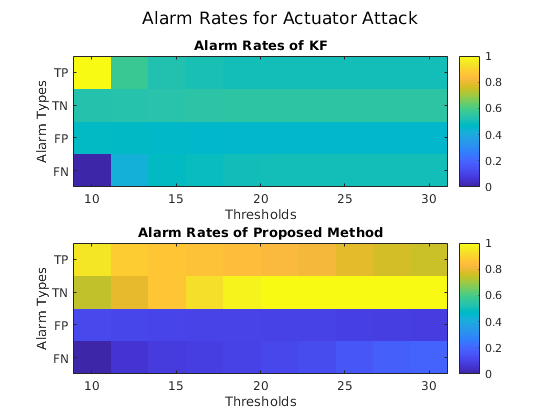}
    \caption{The metrics of different detectors are defined as follows: the true positive rate (TP), the true negative rate (TN), the false positive rate (FP), and the false negative rate (FN).}
    \label{fig:metric}
\end{figure}

As shown in Figure \ref{fig:metric}, despite the threshold changes from 10 to 25, the true positive alarm rate and true negative alarm rate of the proposed method are significantly higher than those of the Kalman filter when using the same threshold.

\begin{figure}[h]
    \centering
    \includegraphics[clip, trim=0.0cm 0.0cm 0.0cm 0.0cm, width=1\linewidth]{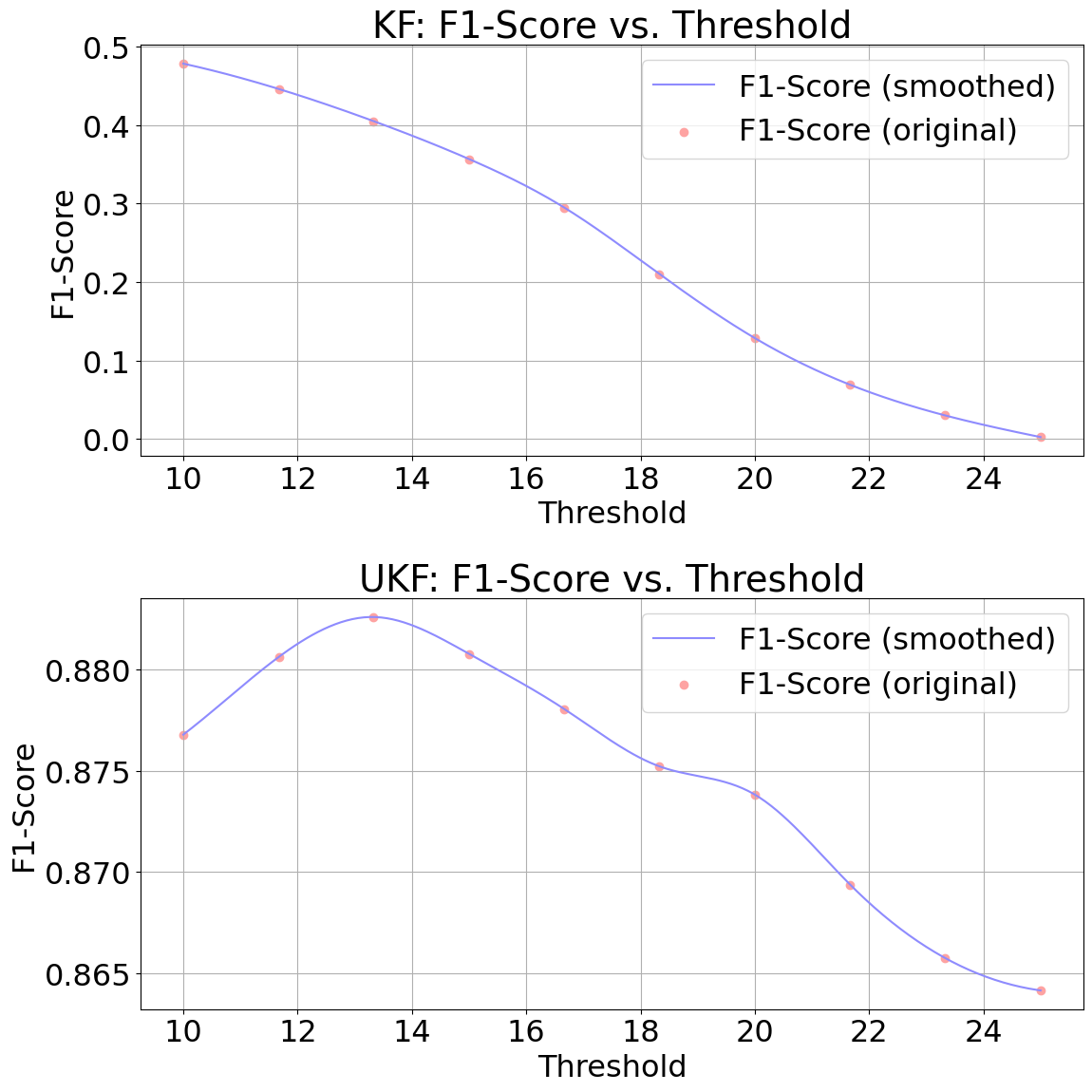}
    \caption{F1 Score of Kalman Filter and the Proposed Method.}
    \label{fig:f1-score}
\end{figure}

To achieve a more precise quantification of the research data, we utilized the F1-score \cite{bishop2006pattern}. As shown in Figure~\ref{fig:f1-score}, the F1-score of the Kalman filter shows a decreasing trend, while the F1-score of the \ac{ukf} first increases and then decreases. Based on this pattern, we selected the appropriate thresholds for the Kalman filter and the proposed method. The thresholds are 10 for the Kalman filter and 13.33 for the proposed method.

\begin{table}[h] 

\caption{Optimal Thresholds and Confusion Matrix Values for Kalman Filter and the Proposed Method}
\centering
\label{tab:optimal_thresholds}
\begin{tabular}{|c|c|c|} 
    \hline
    \textbf{Metric} & \textbf{Kalman filter} & \textbf{Ours} \\ \hline
    Optimal Threshold & 10.0 & 13.33 \\ \hline
    F1-Score & 0.4784 & 0.8826 \\ \hline
    \cellcolor{gray!25}
    True Positives (TP) & 42.88\% & \cellcolor{green!25}93.58\% \\ \hline
    \cellcolor{gray!25}
    False Positives (FP) & 36.41\% & \cellcolor{green!25}18.48\% \\ \hline
    \cellcolor{gray!25}
    True Negatives (TN) & 63.59\% & \cellcolor{green!25}81.52\% \\ \hline
    \cellcolor{gray!25}
    False Negatives (FN) & 57.12\% & \cellcolor{green!25}6.42\% \\ \hline
\end{tabular}
\end{table}

Table~\ref{tab:optimal_thresholds} presents the optimal thresholds and corresponding values for the Kalman filter and the proposed method. In this table, all metrics of the proposed method significantly outperform those of the Kalman filter.

\subsubsection{Sensor Attack}

\begin{figure}[h]
    \centering
    \includegraphics[clip, trim=0.0cm 0.0cm 0.0cm 0.0cm, width=1\linewidth]{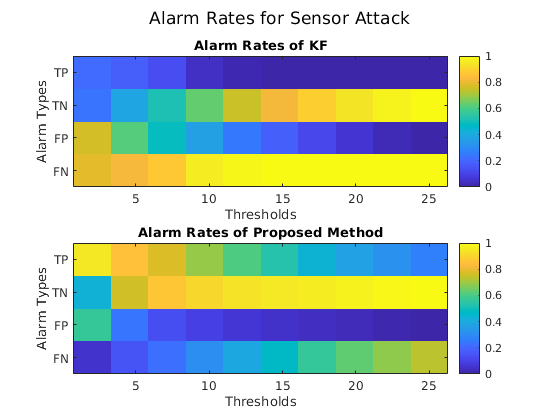}

    \caption{The metrics of different detectors are defined as follows: the true positive rate (TP), the true negative rate (TN), the false positive rate (FP), and the false negative rate (FN).}
    \label{fig:metric2}
\end{figure}
We obtained similar results for \ac{ad} on the sensor, as shown in Figure~\ref{fig:metric2}. Both the true positive and true negative rates of the proposed UKF method are significantly higher than those of the Kalman filter when choosing the best F1-Score.

\section{Conclusion}
In conclusion, we proposed a novel framework for attack detection in vehicles with unknown systems. We exploited a learning-based model to predict and compare observations against expected behavior.  Therefore, compared to the Kalman filter, the proposed approach based on the \ac{ukf} is capable of detecting \ac{dos} attacks from the sensor and actuator without prior knowledge of the system parameters. Accordingly, by exploiting \ac{ukf}'s capabilities in handling nonlinearity, our proposed algorithm demonstrated a significant advantage over the traditional Kalman filter for detecting \ac{dos} attacks on sensors and actuators. In detail, through extensive simulations of our proposed algorithm, we observed that our method outperforms the Kalman filter by demonstrating substantial results in both true positive alarm rate and true negative alarm rate. Enhancing the filtering design for vehicle incursion detection can be the primary focus of future research. By precisely simulating intricate dynamics, investigating robust particle filters may improve anomaly identification even further and strengthen vehicle security. Moreover, the presented framework can be extended to other types of attacks such as \ac{fdi}.

\section{Acknowledgement}
We gratefully acknowledge the financial support provided by BlackBerry and the Natural Sciences and Engineering Research Council of Canada (NSERC). Furthermore, we acknowledge BlackBerry QNX's assistance and cooperation with this study. Their assistance has been crucial to the accomplishment of this task.
\bibliographystyle{apalike}
{\small
\bibliography{main}}

\end{document}